\pdfoutput=0
\documentclass[aps,twocolumn,prl,superscriptaddress,reprint]{revtex4-2}
\usepackage{graphicx,color}  
\usepackage{dcolumn}   
\usepackage{bm}        

\usepackage{epsfig,dsfont,amssymb,amsmath,amsthm,amsfonts,amsbsy,mathtools,mathrsfs,soul}

\definecolor{red}{rgb}{1,0,0}
\definecolor{blue}{rgb}{0,0,1}
\definecolor{black}{rgb}{0,0,0}

\begin{document}
\title{Criticality in the Luria-Delbr\"uck model with an arbitrary mutation rate}

\author{Deng Pan}
\thanks{D.P. and J.L. contributed equally to this work.}
\affiliation{John A. Paulson School of Engineering and Applied Sciences, Harvard University, Cambridge, MA 02138, USA}

\author{Jie Lin}
\thanks{D.P. and J.L. contributed equally to this work.}
\affiliation{Center for Quantitative Biology, Peking University, Beijing 100871, China}

\author{Ariel Amir}
\email{ariel.amir@weizmann.ac.il }
\affiliation{Department of Physics of Complex Systems, Faculty of Physics, The Weizmann Institute of Science, Rehovot 7610001, Israel}

\date{\today}
	
\begin{abstract}
The Luria-Delbr\"uck model is a classic model of population dynamics with random mutations, that has been used historically to prove that random mutations drive evolution. In typical scenarios, the relevant mutation rate is exceedingly small, and mutants are counted only at the final time point. Here, inspired by recent experiments on DNA repair, we study a mathematical model that is formally equivalent to the Luria-Delbr\"uck setup, with the repair rate $p$ playing the role of mutation rate, albeit taking on large values, of order unity per cell division. We find that although at large times the fraction of repaired cells approaches one, the variance of the number of repaired cells undergoes a phase transition: when $p>1/2$ the variance decreases with time, but, intriguingly, for $p<1/2$ even though the fraction of repaired cells approaches 1, the variance in number of repaired cells \emph{increases} with time. Analyzing DNA-repair experiments, we find that in order to explain the data the model should also take into account the probability of a successful repair process once it is initiated. Taken together, our work shows how the study of variability can lead to surprising phase-transitions as well as provide biological insights into the process of DNA-repair.
\end{abstract}
\maketitle

The Luria-Delbr\"uck experiment \cite{luria1943mutations} is a remarkable example where the analysis of random fluctuations leads to deep insights. A population of bacteria was grown, from an initial population of $~100$ to about $10^9$ cells, which are then exposed to a virus (i.e., a bacteriophage). The number of survivors is counted (by plating and counting colonies), which is the main output of the experiment. In a large fraction of cases, the result is zero -- no bacteria survive the viral attack. But in some instances, \textit{hundreds} of bacteria survive. While many scientists would toss away such non-reproducible experiments, Luria and Delbr\"uck realized that the large variance:mean ratio they observed is itself the key experimental result, that rules out the Lamarckian picture of adaptation and was consistent with random mutations (that lead to viral resistance) occurring during the population growth. 

This seminal work was later shown to have a profound mathematical structure. While in the original paper, only the first and second moments of the distribution were evaluated (and shown consistent with the experiments), the full distribution was studied in later works \cite{lea1949distribution}. It is also referred to as the ``jackpot distribution" (since it has a heavy tail scaling as $1/x^2$, resulting from the rare events where a bacterium ``hits the jackpot" and acquires a mutation early on in the lineage tree), and in a certain limit it approaches the Landau distribution \cite{landau1944energy}, which Lev Landau studied in the context of the energy distribution of fast particles colliding and ionizing molecules in their path. Mandelbrot also showed an intriguing connection to the Levy-stable distribution and the generalized central-limit-theorem \cite{mandelbrot1974population}. See Ref. \cite{zheng1999progress} for a review, and Ref. \cite{kessler2013large} for further mathematical progress. In one commonly used model, growth is essentially deterministic and each cell cycle takes an identical, precise duration. Hence, the only source of stochasticity is related to the occurrence of random mutations. This model was used in the original work, due to its conceptual simplicity and analytical advantage.  We will refer to this model as the synchronous growth model, since in this case, all cells of a given generation divide at precisely the same time (note that we are also assuming that mutants and wild-type cells have identical doubling time). In another, more realistic, mathematical model, cell divisions are also a source of stochasticity \cite{kessler2013large}, though the qualitative behavior of both models is similar.

Since the original work, in addition to the mathematical advances described above, the LD model and its variants have seen wide applications in different fields. The Luria-Delbr\"uck fluctuation assay is useful in order to determine the relevant mutations rates leading to important phenomena such as antibiotic resistance \cite{Oakberg.1947.Luria}.  
More recently, the LD model has been used to study drug-resistant cancer cells and carcinogenesis \cite{Jr.2012,Bozic.2014}. In a recent line of research, the Luria-Delbr\"uck fluctuation assay is used to assess \textit{phenotypic variability}, i.e. traits that are not genetically encoded and thus not fully heritable, with a memory spanning only a few generations. Nonetheless, the fluctuations can provide valuable information regarding the underlying mechanisms, see Ref. \cite{singh2023probing} for a review.

In a recent experimental setup, yeast cells were genetically engineered such that a segment of their DNA contains repetitive sequences, also known as microsatellites, and upon removal of this segment (guided by CRISPR), cells concurrently begin to express Green Fluorescent Protein (GFP) such that their status may be monitored using video microscopy \cite{vertti2022time}. In this double-strand break repair (DSBR) experiment, the GFP-expressing status is fully heritable, and although the repair is not due to a random mutation, we may consider it as a random process (occurring with some probability $p$ per generation) -- hence mathematically it meets the conditions of the LD model:(1) The occurrence of random mutations leading to a distinct trait (e.g. GFP positive in the yeast experiment)
(2) The heritability of the trait. Nonetheless, this setup is profoundly different than that of the original LD experiment in terms of the relevant parameters: in the LD setup, the mutation probability per generation is of the order of $10^{-9}$, as is clear from the fact that out of the $o(10^9)$ division events, only a handful at most will have relevant mutations. In the DNA-repair experiment, on the other hand, $p$ was found to be of the order of unity \cite{vertti2022time}. This distinction necessitates consideration of the \textit{irreversible} version of the LD model, where a mutated cell cannot revert to its wild-type state. This premise was less critical in the original LD model due to the negligible probability of such reversion with the typically small population of mutated cells. Additionally, in this experiment, the number of repaired cells is tracked \textit{throughout} the experiment, not only at the final time point (as is the case in the original LD setup). 

Here, we analyze the fluctuations observed in a LD setup in this very different regime. Ultimately, all cells will become repaired (no matter what the value of $p$ is). However, we find that the variance in the number of repaired cells exhibits two distinct fates, depending on the value of $p$: it may decay over time, as is perhaps the naive intuition since all cells are eventually repaired. However, for $p<1/2$, we find that the variance \textit{increases exponentially} with time, due to the subpopulation of non-repaired cells (the fraction of which monotonically decreases with time).

Comparing the LD model with the DSBR experimental data, we find that by fitting the repair probability $p$, the model can explain the mean fraction of repaired yeast cells as a function of time, but leads to an incorrect prediction for the standard deviation vs. time, growing slower than the experimental data. To explain the deviation, we introduce a modified LD model, where each repair initiated has two possible outcomes: either successful repair or dormant cells that do not divide. The modified model aligns with the experimental data for both the mean and standard deviation.

\begin{figure}[htb!]
	\includegraphics[width=0.4\textwidth]{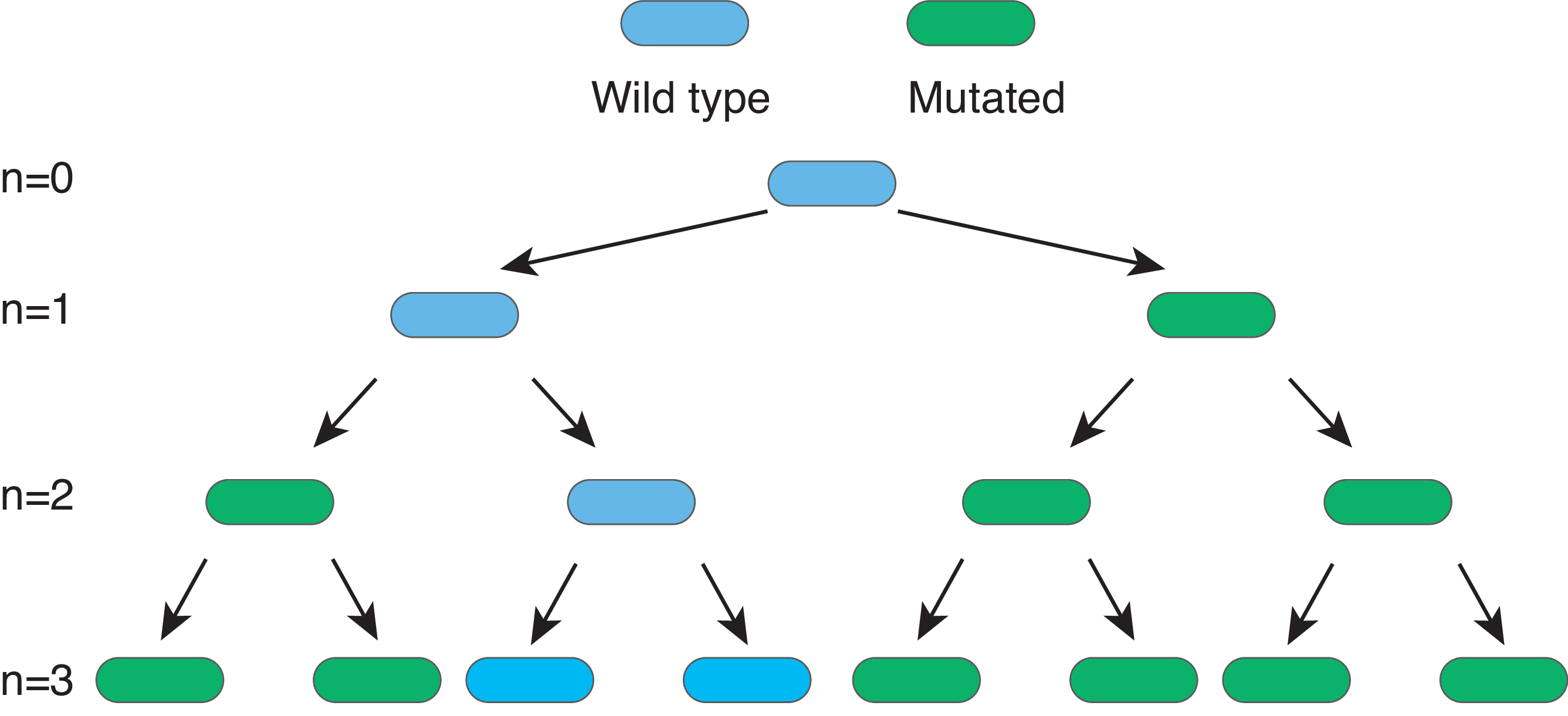}
	\caption{A schematic of the model. Each cell has a constant probability $p$ to mutate per cell cycle, and the offspring of mutated cells inherit the mutation.}
	\label{fig:illus}
\end{figure}

{\it Synchronous growth model.---} We consider an exponentially growing population descending from a single wild-type (WT) cell, where each cell has a constant probability $p$ to mutate \textit{irreversibly} per cell cycle (corresponding to the repair of the microsatellite region of the genome in the DSBR experiment). We assume that the timing of cell division is synchronous so that a population starting from a single WT cell, which we define as generation $0$, will have $2^n$ cells in generation $n$ (Figure \ref{fig:illus}). It is well known that this branching process has a phase transition at $p=1/2$ \cite{KimmelMarek2015BPiB}: above this critical value, at long times \emph{all} of the cells in the population become mutants, while below this critical value, there is a finite probability $P_{f}$ of having WT cells also at arbitrarily long times. It is straightforward to show that $P_{f}$  obeys the following equation, also shown in Figure 2(a) for finite number of generations:
\begin{equation}
	P_{f}=\Big(\frac{p}{1-p}\Big)^2\label{eq:Pf}.
\end{equation}
Inspired by the original LD problem, where the study of variance was key in distinguishing Darwinian and Larmarckian evolution, we will be interested in the \emph{variance} of the number of mutated cells over time. Intriguingly, the temporal dynamics of the variance will also manifest a phase transition at $p=1/2$. Moreover, as we shall see, comparing this property with experimental data will also lead us to useful insights.

To this end, we will first consider a recursive equation for the number of mutated cells over time. 
\begin{equation}
	m_n=2m_{n-1} + \sum_{i=1}^{2(2^{n-1}-m_{n-1})} \xi_i\label{eq:rec1},
\end{equation}
where $\xi_i$ is a random variable, which is $1$ with probability $p$ and 0 with probability $1-p$. The first term on the RHS comes from the doubling of mutated cells in generation $n-1$. The second term comes from non-mutated cells, which may mutate in generation $n$. 

We can compute the average number of mutated cells by applying Wald's equation to decouple the summation bound ($2^n-2m_{n-1})$ and the random variable ($\xi_i$) \cite{wald1944cumulative}, resulting in the following solution
\begin{equation}
	\langle m_n\rangle =2^n(1-(1-p)^n).\label{eq:mmean}
\end{equation}
Eq. (\ref{eq:mmean}) shows that the average number of mutated cells approaches the total number in the long time limit.
\begin{figure}[htb!]
	\includegraphics[width=0.49\textwidth]{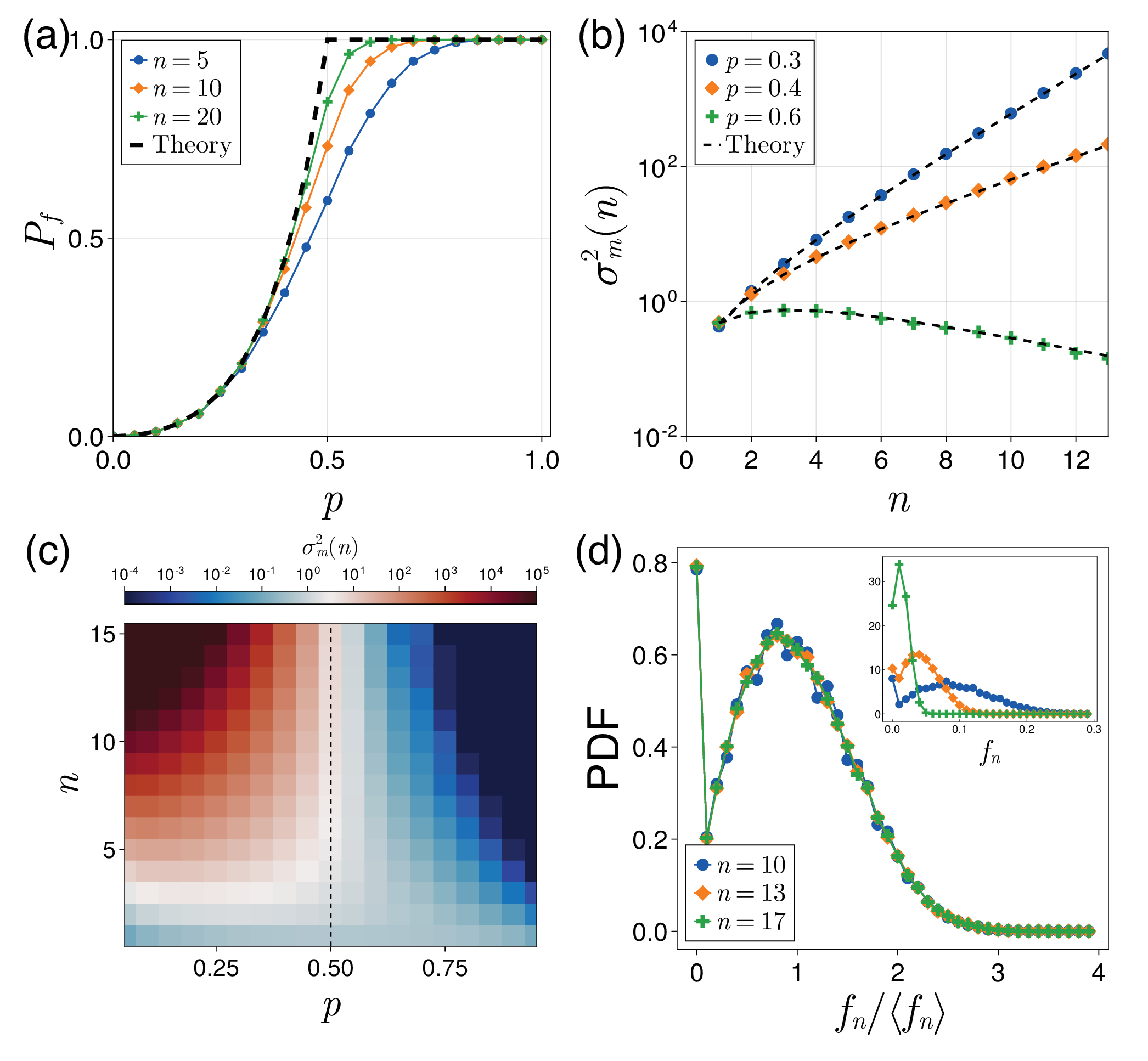}
	\caption{Simulations of the synchronous growth model. (a) The fraction of simulations where all cells are mutated by a given generation. The dashed line is the theoretical prediction in the infinite time limit, see Eq. (\ref{eq:Pf}). (b) The variance of the number of mutated cells as a function of the generation. (c) The variance of the number of mutants $\sigma_m^2$ {\it vs.} both $p$ and $n$. (d) The probability distributions of the fraction of WT cells can collapse to a single distribution when normalized by their mean. The unscaled data are shown in the inset. It combines a delta function at $f_n=0$ (since it is an absorbing fixed point) and a continuous distribution at $f_n>0$.}
	\label{fig:syn}
\end{figure}
We next compute the variance of the number of mutated cells. If we denote the number of wild-type cells as $w_n$, then we have 
\begin{equation}
    w_n + m_n = 2^n.
    \label{eq:totalCellNumberDiscreteLD}
\end{equation}
Therefore the variance of the number of mutated cells should be equal to the variance of wild-type cells $\sigma_m(n) = \sigma_w(n)$. 

By applying the Blackwell–Girshick equation to the recurrence relation, we are able to find the recursive relation of the variance \cite{blackwell1946functions} ,
\begin{equation}
    \sigma^2_w(n) = 4(1-p)^2 \sigma^2_w(n-1) + p(1-p)^n2^n.
\end{equation}
Solving the equation, we obtain:
\begin{equation}
	\sigma^2_m(n)=\frac{p}{1-2p}\{[2(1-p)]^{2n}-[2(1-p)]^n\}.
	\label{eq:sigmam}
\end{equation}

A critical transition happens at $p_c=1/2$, below which the variance diverges in the long time limit $\sigma^2_m(n) \rightarrow \frac{p}{1-2p}[2(1-p)]^{2n}$, and above which the variance vanishes in the long time limit $\sigma^2_m(n) \rightarrow \frac{p}{2p-1}[2(1-p)]^n$ (Figure \ref{fig:syn}b,c). At $p=p_c$, $\sigma^2_m(n)=n/2$ diverges linearly. One can also express the variance as a function of the total cell number using $N=2^n$, which will allow us to compare the predictions of the synchronous growth model with simulations based on asynchronous growing populations:
\begin{equation}
	\sigma^2_m(N)=\frac{p}{1-2p}\left( N^{2+2\frac{\ln 1-p}{\ln 2}}-N^{1+\frac{\ln 1-p}{\ln 2}}\right) .
	\label{eq:sigmamN}
\end{equation}

We define the fraction of WT cells in generation $n$ as $f_n=1-m_n/2^n$ so that it is between $0$ and $1$. From Eq. (\ref{eq:mmean}) and Eq. (\ref{eq:sigmam}), we find that the coefficient of variation (CV, the ratio of the standard deviation and the mean) of $f_n$,
\begin{equation}
\sigma_f^2(n)/\langle f_n\rangle^2 \xrightarrow[n \rightarrow \infty] \quad
\begin{cases}
	p/(1-2p), &\text{when } p < p_c \\
	\infty, &\text{when } p\ge p_c
\end{cases}
\end{equation}

In fact, our numerical simulations suggest a stronger result, namely, that for $p<p_c$ the distribution is approximately scale-invariant (also for small values of $n$), and takes the following form:
 \begin{equation}
 	P(f_n)=\frac{1}{\langle f_n\rangle} H\Big (\frac{f_n}{\langle f_n\rangle}\Big),\label{eq:pfn}
 \end{equation}
see Figure \ref{fig:syn}d. In the Supplementary Materials (SM), we show rigorously that this form is approached asymptotically (for large $n$), using the theory of classical branching processes and generating functions \cite{KimmelMarek2015BPiB}. 

Before comparing these analytical results with experimental data, we also want to verify that they are valid within the context of more realistic models that account for asynchronous growth.

\begin{figure}[htb!]
	\includegraphics[width=0.49\textwidth]{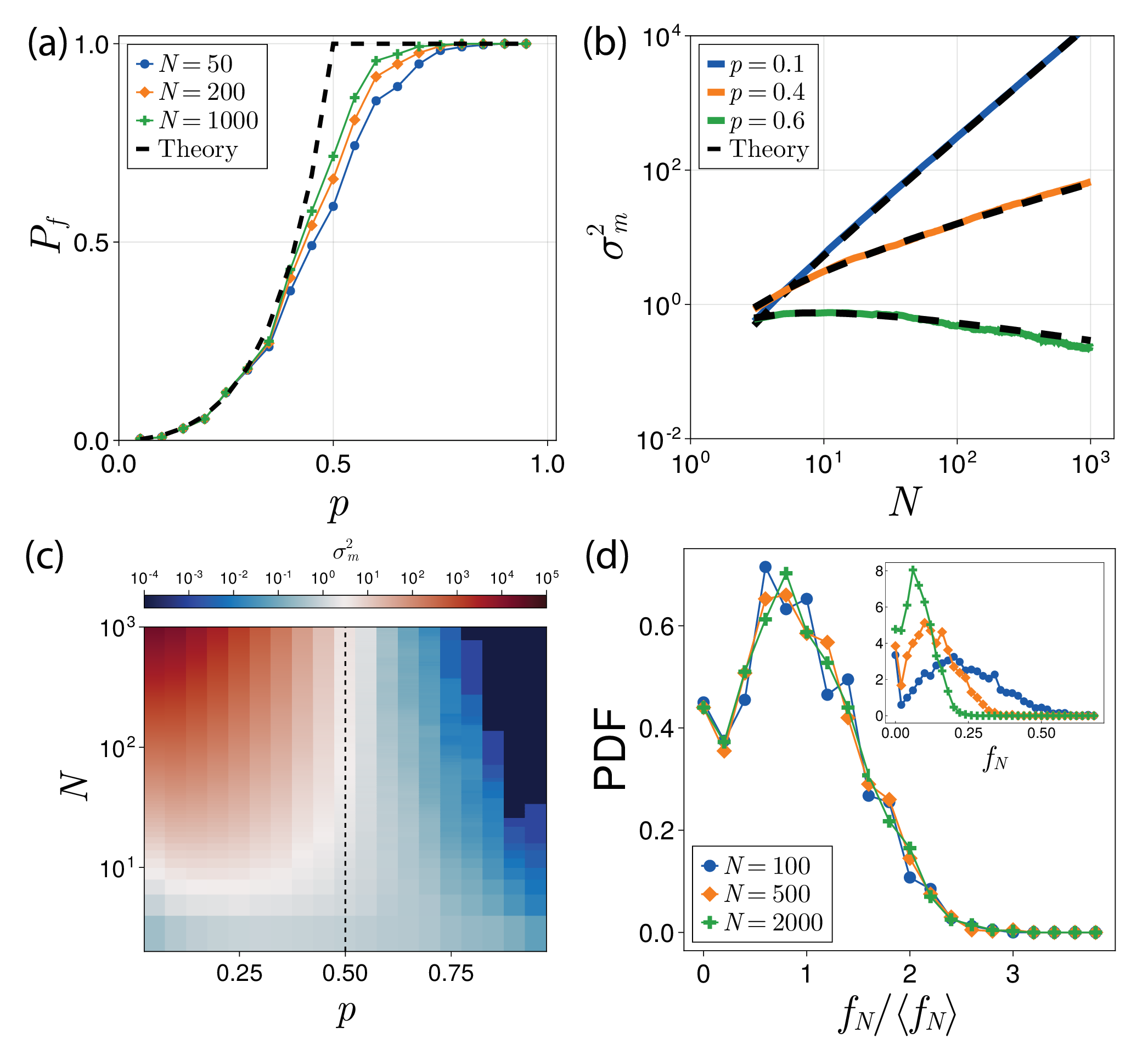}
	\caption{Simulations of the asynchronous growth model. Panels (a)-(d) correspond precisely to Figure 2, albeit with the generation number $n$ replaced by the total number of cells $N$, which we condition upon.}
	\label{fig:rand}
\end{figure}

{\it Asynchronous populations.---} We simulate an asynchronous growing population in which the generation times are \textit{correlated} random variables. To this end, we use a phenomenological model utilized in prior works on microbial growth \cite{cerulus2016noise, Barber.2021,Lin2020, lin2020single}, where the generation time of the daughter cell is related to that of the mother cell:
\begin{equation}
	\ln(t_d)= a \ln(t_m)+b+\xi.
\end{equation}
Here, $a$ and $b$ are constants and $\xi$ is a Gaussian noise. We choose the parameters such that the CV of the generation time distribution is $0.1$. When we plot the simulation results for the variance vs. the number of cells, we find that asynchronous model exhibits a phase transition analogous to the one discussed previously for the synchronous model (Figure \ref{fig:rand}a,b,c).
Note that this is not the case when the variance is plotted vs. time, since the stochastic divisions may lead to growing variance vs. time also for $p>1/2$ (Figure S1). Interestingly, our simulations suggest that also for the asynchronous model, the distribution of the fraction of WT cells (conditioned on the number of cells) is scale-invariant (Figure \ref{fig:rand}d). 

\begin{figure}[htb!]
	\includegraphics[width=0.45\textwidth]{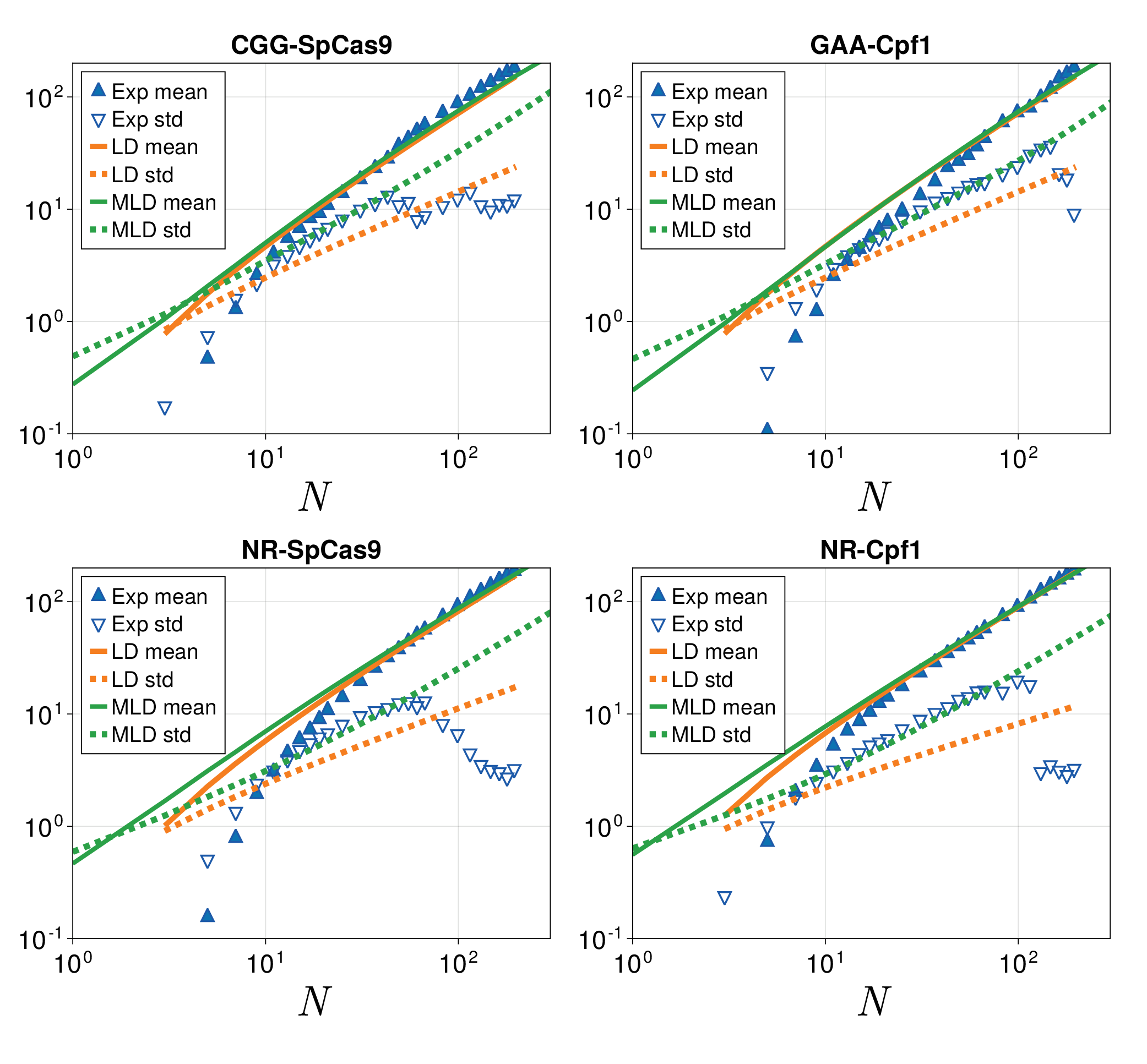}
	\caption{The DSBR experimental data compared with LD model and modified LD model simulations. The fitting parameters are listed in Table. S1.}
	\label{fig:expVsModifiedLD}
\end{figure}

{\it Comparison with experiments.---} Next, we compare our theoretical predictions with the DNA-repair data of Ref. \cite{vertti2022time}. The double-strand break repair experiment was studied for 8 different combinations, including two endonucleases (Cas9 and Cpf1) and four target sequences (NR,CGG,GAA and CTG). The endonucleases, enzymes which can cut DNA strands at specific locations, have different affinities to different target DNA sequences, which therefore could result in different DNA break and repair probabilities. Out of the 8 combinations, 4 of them are identified as high-efficacy error-free repair, which means the DSBR process happened on most of the wells in the microfluidic array and did not significantly affect the cell fitness, hence they fall into the high ``mutation'' rate LD model cases if the DSBR process is thought as a random inheritable and irreversible mutation \cite{vertti2022time}. Therefore, we focus on these for our analysis (for the other conditions, the number of repaired cells is too low). However, they do not agree with the prediction of the LD model in standard deviation and variance when conditioned on the total number of cells $N$, as shown in Figure \ref{fig:expVsModifiedLD}. While by fitting the parameters of the LD model we can capture the \textit{mean} number of repaired yeast cells as a function of time, the standard deviation given by the LD model grows slower than the experimental data (the standard deviation of the experiment is valid until it decreases sharply, which is the result of the decrease in sample size due to experiment time constraint). To explain the deviation, we introduce a modified LD model with a probability of successful repair that can match the experimental data in both the mean and standard deviation. 

{\it Modified LD model.---} In the yeast DSBR experiment, several foundational assumptions of the LD model might not hold, including: (1) mutations not affecting cell fitness, whereas repaired cells in the DSBR could exhibit varied fitness; (2) 
for the DSBR setup, once a DNA break occurs, the repair process might have a non-negligible probability to fail; This implies that there are \textit{two} probabilities characterizing each cell cycle (of a break, and of successful repair), in contrast to the LD model where one mutation probability per cell cycle defines the model fully (3) the LD model assumes a constant cell division time, whereas yeast in the DSBR may experience \textit{asymmetric} budding with variable division times. Although any of the three factors could change the population dynamics significantly, one observation of the experiments implies that the most likely cause of the deviation could be the second: for each of the experimental conditions, a large portion (about 30\%) of the wells do not have any repaired cells by the end of the experiment. It is plausible that the broken DNA strands in these wells remain unrepaired. Since cells with broken DNA cannot divide, a modified LD model could introduce repair success probability for the breaking and repair processes so that once a cell is broken it can be either repaired or dormant. The modified LD model shows excellent agreement with the experimental data in both the mean and standard deviation, as shown in Figure \ref{fig:expVsModifiedLD}. It is not a simple result from modeling with more parameters, as we also tested models relaxing assumptions (1) and (3), by introducing different growth rate of repaired cells and adding extra noise to division time, neither alone could make the model fit better with the DSBR experimental data. In addition, the modified LD model with proper fitting parameters can match the distribution of repaired cells across wells, as supported by hypothesis testing (see SM and Table. S1). There is some discrepancy between the model prediction and experimental data in the initial phase when the cell count is low. Firstly, the figure is on a log-log scale, so the deviation is not large in the absolute magnitude. Secondly, the deviation can be attributed to a delay of GFP expression following DNA repair, which could span up to 6 hours (equivalent to approximately 1-2 cell generation periods) \cite{vertti2022time}. This delay results in a reduced count of GFP-positive cells in the early stages.

{\it Discussion.---} In this work, we studied the Luria-Delbr\"uck model with a large mutation rate, which is relevant to the DNA-repair experiment. We found that the variance of the number of mutated cells exhibits a phase transition at $p=1/2$, below which the variance increases with time, and above which the variance decreases with time. We also found that the distribution of the fraction of WT cells is scale-invariant when $p<p_c$. Both are verified by synchronous and asynchronous LD model simulations. We compared the LD model with the DSBR experimental data and found that by introducing a probability of successful repair, a modified LD model can match the experimental data in both the mean and standard deviation. Our results demonstrate that by studying the fluctuations of the temporal dynamics across different wells, we can learn about the underlying mechanisms of the DNA repair process. For example, within the four examined combinations of endonucleases and target sequences, the successful repair probability of non-repeated control sequences (NR) is found to be twice as high as that for CGG and GAA sequences. This difference is not discernible through the average number of repaired cells (see Table S1). 
In fact, the inference of new insights from the \textit{variability} rather than the \textit{mean} behavior, via the utilization of mathematical and physical models, is exemplified in various other examples in biology \cite{Amir2018LearningFN}. These include the original Luria-Delbr\"uck problem, as well as, more recently, in studies interrogating the microbial cell cycle \cite{Ho2018ModelingCS, Kar2023UsingCI}, the mammalian cell cycle \cite{Sandler2015LineageCO} and bacterial populations in droplets \cite{taylor2022tracking}, to name but a few.

{\it Acknowledgement.---} We would like to thank Ethan Levien and Prathitha Kar for their invaluable discussions and insights that significantly contributed to this work.

%
\end{document}


\newcounter{suppequation}
\title{Criticality in the Luria-Delbr\"uck Model with an arbitrary mutation rate}

\author{Deng Pan}
\thanks{D.P. and J.L. contributed equally to this work.}
\affiliation{John A. Paulson School of Engineering and Applied Sciences, Harvard University, Cambridge, MA 02138, USA}
\author{Jie Lin}
\thanks{D.P. and J.L. contributed equally to this work.}
\affiliation{Center for Quantitative Biology, Peking University, Beijing 100871, China}

\author{Ariel Amir}
\affiliation{Department of Complex Systems, Faculty of Physics, The Weizmann Institute of Science, Rehovot 7610001, Israel}
\email{ariel.amir@weizmann.ac.il }
\date{\today}
\let\oldtheequation\theequation
\let\oldalign\align
\let\endoldalign\endalign
\let\oldtheHequation\theHequation 
\let\oldequation\equation
\let\endoldequation\endequation

\renewenvironment{equation}
  {\stepcounter{suppequation}\renewcommand{\theequation}{S\arabic{suppequation}}\oldequation}
  {\endoldequation}
\renewenvironment{align}
  {\stepcounter{suppequation}\renewcommand{\theequation}{S\arabic{suppequation}}\oldalign}
  {\endoldalign}
\renewcommand{\thefigure}{S\arabic{figure}}

\renewcommand{\thetable}{S\arabic{table}}
\maketitle
\section{LD model is a special case of branching process}
 LD model can be regarded as a special case of traditional branching processes (specifically uncorrelated asynchronous LD model is a multitype age-dependent branching process). Theory of branching processes and its application in biology are documented in Ref. \cite{kimmel2015branching}. 

To show that our results are consistent with the theory of branching processes, first, we will show that the discrete LD model can be mapped to a simple Galton-Watson process. In the discrete LD model, the total number of cells is fixed at $n$th generation (Eq.\ref{eq:totalCellNumberDiscreteLD}), therefore knowing the mean and variance of wild-type cells is equivalent to knowing the mean and variance of the mutant cells. Since the mutation is irreversible, the distribution of wild-type cells is exactly the same as that in a Galton-Watson process with probability $(1-p)^2$ of having two daughter cells and $2(1-p)p$ of having one daughter cell. According to the theory of Galton-Watson processes, the probability generating function(pgf) for one generation is
\begin{equation}
    g_1(s) = p^2 + 2(1-p)ps + (1-p)^2 s^2 = \Big ( p+(1-p)s \Big )^2,
\end{equation}
which iterates to the pgf for $n$th generation
\begin{equation}
    g_n(s) = g_1[g_{n-1}(s)].
\end{equation}
The mean of wild-type cells can be calculated through the first derivative of the pgf:
\begin{equation}
    \langle w_n \rangle = g'_n(s)\Bigr|_{s=1} = 2^n (1-p)^n.
\end{equation}
The factorial moment of the distribution can be calculated as well:
\begin{equation}
    E[w_n(w_n-1)(w_{n}-2)...(w_n-r+1)] = g^{(r)}_n(s)\Bigr|_{s=1}.
\end{equation}
To find the variance of the distribution, we can solve the second-order factorial moment $\mu_n$:
\begin{equation}
    \mu_n = E[w_n^2 - w_n] = g^{(2)}_n(s)\Bigr|_{s=1},
\end{equation}
which leads to the recursive relation by the chain rule of derivatives
\begin{equation}
    \mu_n = 2^{(2n-1)}(1-p)^{2n}+2(1-p)\mu_{n-1}.
\end{equation}
The solution of the recurrence equation is 
\begin{equation}
    \mu_n = \frac{(p-1) \left(2^n (1-p)^n-1\right) (2-2 p)^n}{2 p-1}.
\end{equation}
Therefore the variance can be directly calculated:
\begin{align}
    \sigma_w^2(n) &= \mu_n + \langle w_n \rangle -\langle w_n \rangle^2 \nonumber  \\
    & = \frac{p}{1-2p}\{[2(1-p)]^{2n}-[2(1-p)]^n\},
\end{align}
which is consistent with previous results Eq. (\ref{eq:sigmam}). 
It has been proved that in the Galton-Waston process, the distribution of $w_n/\langle w_n \rangle$ will converge in the supercritical case ($p<1/2$):
\begin{equation}
    \lim_{n\rightarrow\infty} \frac{w_n}{\langle w_n \rangle} = W, \text{with prob } 1,
\end{equation}
which is equivalent to the statement in the main text Eq. (\ref{eq:pfn}). 
When simple asynchronous division is taken into consideration (i.e. the division time of every cell is independent), we can get similar results from the classical Bellman-Harris model (Ref. \cite{kimmel2015branching}).  

\section{Model fitting methods}
To fit the experiment with the original LD model, in which we treat the ``mutation" rate as equivalent to the break and repair rate of the DSBR experiment, we simply fit the mean repaired cells with the analytical solution of the LD model (main text Eq. (\ref{eq:mmean})) to get the ``mutation" rate using the nonlinear least square method. Then we can use the fitting results to plot the standard deviation of the repaired cells (main text Eq. (\ref{eq:sigmamN})). We then use the best-fit ``mutation" rate to do the Kolmogorov-Smirnov (K-S) test with 10,000 simulations and the experimental data at the end of the experiment when the standard deviation starts to decrease. The K-S test is a nonparametric statistical test that evaluates the degree of similarity between two sample distributions or a sample distribution and a reference probability distribution. It does so by comparing the cumulative distribution functions (CDFs) of the two distributions, calculating the maximum distance between these CDFs as its test statistic. The resulting p-value indicates the likelihood of observing the calculated test statistic under the null hypothesis, which posits no significant difference between the compared distributions. A p-value greater than a chosen significance level (commonly 0.05) suggests that the test fails to reject the null hypothesis, implying that the distributions are not statistically distinguishable based on the data analyzed.

Fitting the experiment with the modified LD model is more complicated, since we do not have an analytical solution when the probability of successful repair is not equal to 1. We use the following method to fit the experiment. First, we run a parameter scan of break probability $p_b$ and repair probability $p_r$, both between 0 and 1 with step size 0.01. At each step, we run 10,000 simulations of the same parameters and then calculate its mean and standard deviation over time and binned by the cell count. Then we measure the distance between the simulation results and the experimental data by summing the square of the difference between the simulation and experimental data (including both mean and standard deviation). The best-fit parameters are the ones that minimize the distance. Lastly, we use the best-fit parameters to do the K-S test with the experimental data at the end of the experiment.

The fitting results are shown in Table \ref{tabS:MLDParameters}. A priori, one might think that the modified LD model maps to the original one with an effective mutation rate equaling the \textit{product} $p_b\times p_r$. Indeed, we find that the best fit values of $p_b$, $p_r$ roughly obey this relation (see Table S1). However, we know that no rigorous mapping between the two models may be possible, since only the modified LD model can simultaneously capture the dynamics of both mean and variance. In addition, our K-S tests suggest a stronger result that the distribution of repaired cells is aligned with the MLD model while the LD model fails to match the population distribution.


\newpage
\begin{table}[]
\begin{tabular}{|l|c|c|c|c|c|c|}
\hline
         & \multicolumn{1}{l|}{$p_b$} & \multicolumn{1}{l|}{$p_r$} & \multicolumn{1}{l|}{$p_b\times p_r$} & \multicolumn{1}{l|}{LD model ``mutation" rate} & \multicolumn{1}{l|}{K-S test p-value for MLD} & \multicolumn{1}{l|}{K-S test p-value for LD} \\ \hline
CGG-Cas9 & 0.43                       & 0.32                       & 0.14                                 & 0.17                                          & 0.41                                          & $<10^{-4}$                                   \\ \hline
GAA-Cpf1 & 0.36                       & 0.34                       & 0.12                                 & 0.17                                          & 0.14                                          & 0.021                                        \\ \hline
NR-Cas9  & 0.46                       & 0.50                       & 0.23                                 & 0.23                                          & 0.23                                          & $<10^{-11}$                                  \\ \hline
NR-Cpf1  & 0.50                       & 0.62                       & 0.31                                 & 0.28                                          & 0.28                                          & $<10^{-10}$                                  \\ \hline
\end{tabular}
\captionsetup{justification=raggedright,singlelinecheck=false}
\caption{The best-fit simulation parameters of the original and modified LD model for each experiment, and the Kolmogorov-Smirnov test p-value between the experimental data and the LD/MLD simulation results at the end of the experiment.}
\label{tabS:MLDParameters}
\end{table}

\newpage 
\begin{figure}[h]
	\includegraphics[width=0.6\textwidth]{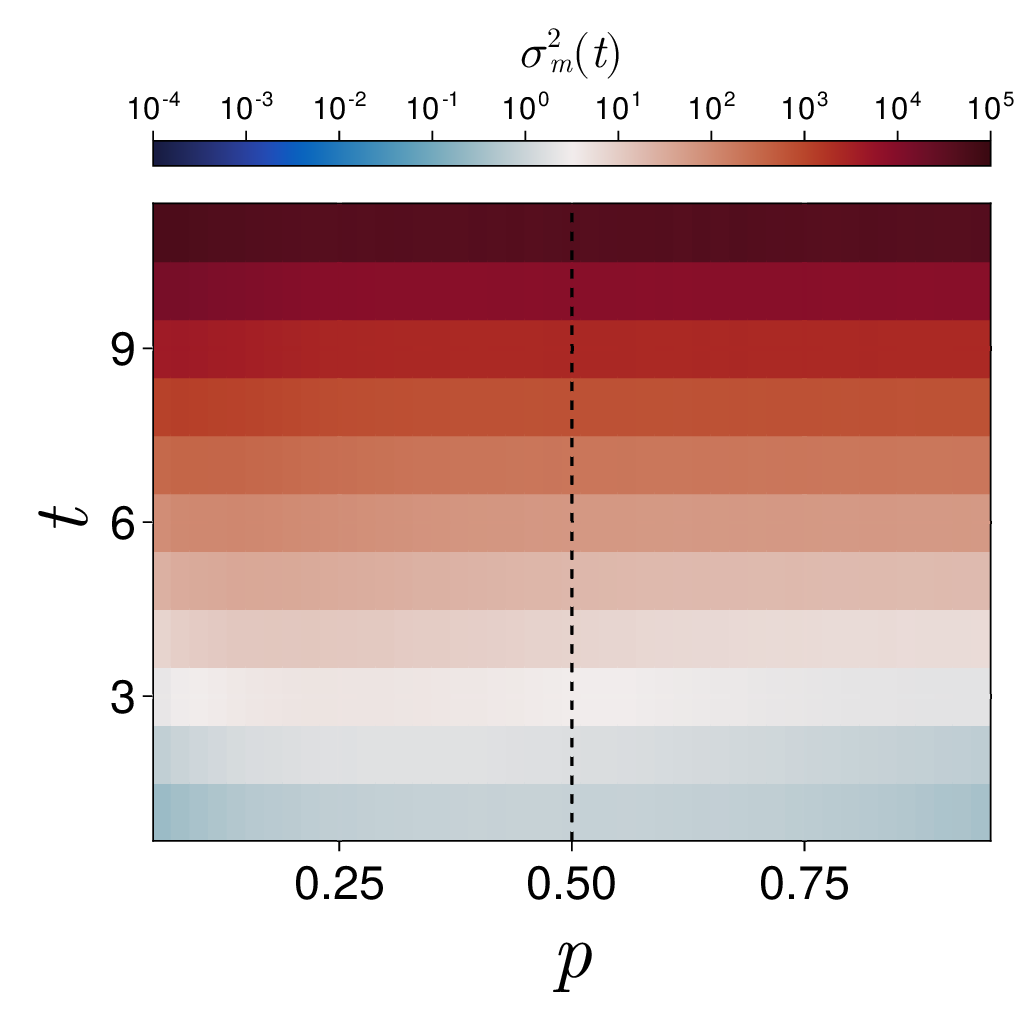}
        \captionsetup{justification=raggedright,singlelinecheck=false}
	\caption{The variance of the number of mutants $\sigma_m^2$ vs. both $p$ and $t$ for asynchronous LD model, showing that the variance always grows with time when it is not conditioned on the cell count. }
\end{figure}
\let\align\oldalign
\let\endalign\endoldalign
\let\theequation\oldtheequation
\let\theHequation\oldtheHequation 
%
\makeatletter\@input{xx.tex}\makeatother